\documentclass[twocolumn,showpacs,superscriptaddress]{revtex4}
\usepackage{amssymb}
\usepackage{amsmath}
\usepackage{graphicx}
\usepackage{epsfig}
\usepackage{amsfonts}

\begin{document}

\title{ Creation of Entanglement between Two Electron Spins \\
Induced by Many Spin Ensemble Excitations }
\author{Qing Ai}
\affiliation{Department of Physics, Tsinghua University, Beijing 100084, China}
\author{Yong Li}
\affiliation{Department of Physics, University of Basel, Klingelbergstrasse 82, 4056
Basel, Switzerland}
\author{Guilu Long}
\affiliation{Department of Physics, Tsinghua University, Beijing 100084, China}
\affiliation{Tsinghua National Laboratory for Information Science and Technology, Beijing
100084, China}
\author{C. P. Sun}
\affiliation{Institute of Theoretical Physics, Chinese Academy of
Sciences, Beijing, 100080, China}
\email{suncp@itp.ac.cn}
\homepage{http://www.itp.ac.cn/~suncp}

\begin{abstract}
We theoretically explore the possibility of creating spin entanglement by
simultaneously coupling two electronic spins to a nuclear ensemble. By
microscopically modeling the spin ensemble with a single mode boson field,
we use the time-dependent Fr\"{o}hlich transformation (TDFT) method
developed most recently [Yong Li, C. Bruder, and C. P. Sun, Phys. Rev. A
\textbf{75}, 032302 (2007)] to calculate the effective coupling between the
two spins. Our investigation shows that the total system realizes a solid
state based architecture for cavity QED. Exchanging such kind effective
boson in a virtual process can result in an effective interaction between
two spins. It is discovered that a maximum entangled state can be obtained
when the velocity of the electrons matches the initial distance between them
in a suitable way. Moreover, we also study how the number of collective
excitations influences the entanglement. It is shown that the larger the
number of excitation is, the less the two spins entangle each other.
\end{abstract}

\maketitle

\section{Introduction}

Since Shor and Grover algorithms \cite{Shor, Grover} were proposed with
various following significant developments, e.g., \cite{long01}, quantum
computing has been displaying its more and more amazing charm against
classical computing. As more progress has been made in this area, it is
urgent to discover various robust, controllable and scalable two-level
systems - qubits as the basic elements for the future architecture of
quantum computers. Generally speaking, electron spins are a natural qubit,
especially the single electron spin confined in a quantum dot for its well
separation and easy addressability. In Ref. \cite{Daniel-DiVincenzo},
electron spins in quantum dots were employed as qubits and two-qubit
operations were performed by pulsing the electrostatic barrier between
neighboring spins. Thereafter, Kane's model made use of the nuclear spins of
$^{31}$P donor impurities in silicon as qubits \cite{Kane}. It combined the
long decoherence time of nuclear spins and the advantage of the well
developed modern semiconductor industry.

In practice, it seems difficult to control the coupling between qubits
because the coupling is based on the overlap of two adjacent spin wave
functions \cite{Daniel-DiVincenzo,Kane}; the coupling is given to be fixed
once photolithography of the chip has been finished. Another feasible way to
induce the controllable inter-spin interaction is to couple two spins by
spin-orbit interaction \cite{Zhao}. However, because of the weakness of
spin-orbit coupling, it is very crucial to find a scheme to manipulate two
spin coupling in the strong interaction regime. In the present paper, we
consider the possibility of creating quantum entanglement of two electron
spins by making them pass through a 2D quantum well containing many "cooled"
nuclear spins (see Fig. \ref{Sketch}). It was discovered that the effective
coupling intensity is increased by a factor of $\sqrt{N}$ when an electron
spin is coupled with an ensemble of $N$ nuclear spins \cite{Taylor}. Such an
electron spin coupled to the nuclei has been considered for cooling the
nuclear ensemble \cite{Taylor2}. Moreover, people have proposed a quantum
computing scheme using a scanning tunneling microscopy with a moving tip as
a commuter to perform the control-not gate between two qubits on the silicon
surface \cite{long02,Berman}. Here, the tip plays the role of the quantum
data bus to coherently link the qubits.

Enlightened by these works, we suggest a new scheme to entangle two electron
spins by a tip since it can be modeled as an ensemble of many spins \cite{Ai}%
. Indeed, when the couplings of the spin to the nuclear ensemble are
quasi-homogeneous, the interaction between the electron spin and the
collective excitation of nuclear spins can be well described in terms of
artificial cavity QED \cite{Song}. Here, the collective excitation can
behave as a single mode boson to realize a quantum data bus, while the
electron spin acts as a two-level artificial atom. With the frequency
selection due to the resonance effect, there is only one mode of collective
excitations interacting with the two spins. Especially, when the Zeeman
splits of all nuclear spins are the same, the single mode excitation can
decouple with other modes \cite{Song}. Then, the coupling system with two
qubit spins and nuclear ensemble just acts as a typical cavity QED system or
spin-boson system. To coherently manipulate the indirect interaction between
the two spins, which is induced by the above mentioned collective
excitation, we need to let two electrons go through the quantum well to
realize a two qubit logical gate operation. Since the moving of electrons
leads to a time-dependent coupling, we need to use some new method to derive
the effective Hamiltonian for the inter-spin coupling. Fortunately, a recent
paper suggested such a time-dependent approach \cite{Li}.

The rest of our paper is organized as follows. In Sec.II, in the low
excitation limit, we simplify the total system we considered above
as two spins interacting with a single mode of the collective
excitation of the nuclei, which forms a cavity-QED under the
quasihomogeneous condition. In Sec. III, we derive the effective
Hamiltonian between the two electron spins by the time-dependent
Fr\"{o}hlich transformation (TDFT) method developed recently in
Ref.\cite{Li}. We remark that the TDFT method can be used to derive
an effective Hamiltonian for a class of cavity QED systems with
time-dependent perturbations. Here, we use this transformation for
the case with time-dependent couplings of two spins to a many spin
ensemble. Section IV contains the discussion of entanglement induced
by the effective Hamiltonian and collective excitation's effect on
the entanglement. In Sec. V we review most of the significant
results. Finally, technical details are given in Appendices
\ref{app:appendix1}-\ref{app:appendix2}.

\section{Model Description}

\begin{figure}[ptb]
\includegraphics[bb=200 356 411 516,width=7 cm]{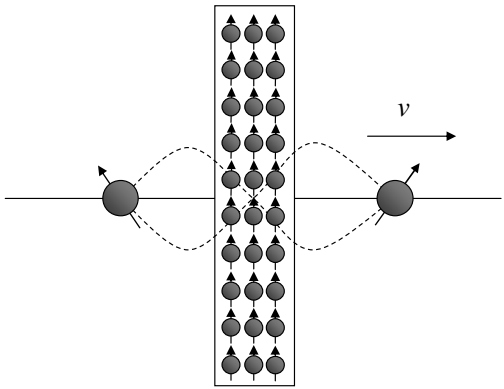}
\caption{Schematic diagram of two electron spins going through a quantum
well consisting of nuclear spins. Two electrons initially located at $%
z_0^{(1,2)}$ move with a uniform speed $v$.}
\label{Sketch}
\end{figure}

We consider a system illustrated in Fig.\ref{Sketch}. Two electrons go
through a quantum well one after the other. The electron spins described by
Gaussian packets with width $a/2$ are initially located at $z_0^{(i)}$ ($%
i=1,2$) and move along the z-direction with a uniform speed $v$. The 2D
quantum well consists of many polarized nuclear spins located in position $%
(x_j,y_j,z_j)$ with $\left\vert x_i\right\vert\leq a$, $\left\vert
y_i\right\vert\leq a$, $\left\vert z_i\right\vert\leq a/5$. When a static
magnetic field is applied to the total system, the Hamiltonian reads as
\begin{align}
H^{S} =&\Omega _{z}(S_{z}^{(1)}+S_{z}^{(2)})+\omega _{z}\sum_{i}I_{z}^{(i)}
\notag \\
&+S_{z}^{(1)}
\sum_{i}g_{1}^{(i)}I_{z}^{(i)}+S_{z}^{(2)}\sum_{i}g_{2}^{(i)}I_{z}^{(i)}
\notag \\
&+S_{+}^{(1)}\sum_{i}\frac{g_{1}^{(i)}}{2}I_{-}^{(i)}+S_{+}^{(2)}\sum_{i}%
\frac{g_{2}^{(i)}}{2}I_{-}^{(i)}+h.c.\text{,}  \label{originalH}
\end{align}
where $S_{z}^{(l)}$ and $S_{\pm }^{(l)}$ ($=S_{x}^{(l)}\pm iS_{y}^{(l)}$) ($%
l=1,2$) are the spin operators for the $l$'th electron spin, $I_{z}^{(j)}$
and $I_{\pm }^{(j)}$ ($=I_{x}^{(j)}\pm iI_{y}^{(j)}$) ($j=1,2,\cdots ,N$)
the spin operators for the $j$'th nuclear spin, $g_{l}^{(j)}$ ($l=1,2$, $%
j=1,2,\cdots ,N$) the hyperfine coupling constants between $l$'th electron
and $j$'th nuclear spin. The first and second terms of Hamiltonian (\ref%
{originalH}) are the Zeeman energies for the electron spins and the nuclear
spins respectively, and the terms besides them are the hyperfine interaction
between the electrons and nuclear spins.

In our setup, the nuclear spins are restricted in a flat square box with $%
\left\vert x_{i}\right\vert \leq a$, $\left\vert y_{i}\right\vert \leq a$, $%
\left\vert z_{i}\right\vert \leq a/5$. We have (for the necessary details
please refer to the Appendix \ref{app:appendix1})
\begin{equation}
g_{1}^{(i)}\simeq g_{0}^{(i)}f_{1}(t)\text{, \ }g_{2}^{(i)}\simeq
g_{0}^{(i)}f_{2}(t)\text{.}
\end{equation}

In Ref. \cite{Song}, the collective excitation of an ensemble of polarized
nuclei fixed in a quantum dot was studied. Under the approximately
homogeneous condition the many-particle system behaves as a single-mode
boson interacting with the spin of a single conduction-band electron
confined in this quantum dot. Likewise, we introduce a collective operator
\begin{equation*}
B=\frac{\sum\nolimits_{i=1}^{N}g_{0}^{(i)}I_{-}^{(i)}}{\sqrt{%
2I_{0}\sum_{j}[g_{0}^{(i)}]^{2}}}
\end{equation*}%
and its conjugate $B^{+}$ to depict the collective excitations in the
ensemble of nuclei with spin $I_{0}$ from its polarized initial state
\begin{equation*}
\left\vert G\right\rangle =\prod_{i=1}^{N}\left\vert -I_{0}\right\rangle _{i}
\end{equation*}%
which is the saturated ferromagnetic state of nuclear ensemble. In
our model, the nuclear spins are fixed in GaAs crystal lattice with
$I_{0}=3/2$, $a=4nm$ and the density of nuclei $n_{0}=45.6nm^{-3}$.
For simplicity, we assume the nuclei are located in a simple cubic
lattice. Thus, we have the average
distance between two neighboring nuclear spins $d=0.28nm$, $g_{max}^{2}/%
\overline{g^{2}}\simeq 11.7$, $N=5046$, where $g_{max}$ is the
maximum of $g_0^{(i)}$ and $\overline{g^{2}}=\sum_i
(g_0^{(i)})^2/N$. On condition that the number of excitations in the
system $n\ll NI_{0}\overline{g^{2}}/g_{max}^{2}$, we have
$[B,B^{+}]\rightarrow 1$. In other words, the collective excitation
described by $B$ can behave as a boson mode in the large $N$ limit
with an initial polarization of all spins in ground (spin down)
state.

In addition to the basic mode denoted by $B$ and $B^{+}$, there exist
auxiliary modes
\begin{equation*}
C_{k}=\frac{\sum\nolimits_{i=1}^{N}h_{i}^{[k]}I_{-}^{(i)}}{\sqrt{2I_{0}(\vec{%
h}^{[k]})^{2}}}
\end{equation*}
for $k=1,2,\ldots ,N$. Here,
\begin{equation*}
\vec{h}^{[k]}=(h_{1}^{[k]},h_{2}^{[k]},\ldots ,h_{N}^{[k]})
\end{equation*}
are $N$ orthogonal vectors in $N$-dimensional space
$\mathbf{R^{N}}$, which can be systematically constructed by making
use of the Gram-Schmidt orthogonalization method \cite{Gram-Schmidt}
starting from
\begin{equation*}
\vec{h}^{[1]}=(g_0^{(1)},g_0^{(2)},\ldots ,g_0^{(N)})\in \mathbf{R^{N}}.
\end{equation*}
Therefore, the Hamiltonian (\ref{originalH}) is rewritten as
\begin{align}
H^{S} \simeq &(\Omega _{z}-f_{1}I_{0}\sum_{i}g_{0}^{(i)})S_{z}^{(1)}+(\Omega
_{z}-f_{2}I_{0}\sum_{i}g_{0}^{(i)})S_{z}^{(2)}  \notag \\
&+f_{1}\Omega (S_{+}^{(1)}B+S_{-}^{(1)}B^{+})+f_{2}\Omega
(S_{+}^{(2)}B+S_{-}^{(2)}B^{+})  \notag \\
&+\omega _{z}\sum_{k}C_{k}^{+}C_{k}+\omega _{z}B^{+}B+H_{p}^{S}  \label{Hs}
\end{align}
Here, the effective Rabi frequency
\begin{equation*}
\Omega =\sqrt{\frac{1}{2}I_{0}\sum\nolimits_{i=1}^{N}\left[ g_{0}^{(i)}%
\right] ^{2}}\text{.}
\end{equation*}%
describes the enhanced coupling of the electron spin to the collective
excitations. And the single particle excitation term
\begin{align}
H_{p}^{S}=(S_{z}^{(1)}f_{1}+S_{z}^{(2)}f_{2})
\sum_{i}g_{0}^{(i)}(I_{z}^{(i)}+I_{0})
\end{align}
can be treated as a perturbation term in the low-excitation limit, which
originates from the inhomogeneity of the couplings.

\section{Effective Inter-Spin Coupling Description}

As shown in Hamiltonian (\ref{Hs}), there are only couplings of two electron
spins with the single mode boson respectively. By making use of the
canonical transformation \cite{Li}, we can eliminate the boson operator and
obtain the effective interaction between the two electron spins. Former
research mainly focused on the case where $f_1(t)\Omega$ and $f_2(t)\Omega$
are time independent \cite{Zheng}. However,the time-independent approach may
not work well in practice. Now, because of the motion of the electrons, we
take the time-dependence of interaction into consideration, namely, $%
f_1(t)\Omega$ and $f_2(t)\Omega$ depend on time.

Let us first summarize the main idea of the time-dependent Fr\"{o}hlich
transformation \cite{Li} so that our paper is self consistent for reading.
Generally speaking, Fr\"{o}hlich transformation \cite{Frohlich,Nakajima} is
frequently used in condensed matter physics to obtain effective interaction
between two electrons by exchanging virtual phonons. For a quantum system
described by Hamiltonian $H(t)=H_{0}+H_{1}(t)$, where $H_{0}$ is time
independent and $|H_{0}|\gg |H_{1}(t)|$, we can make a canonical
transformation
\begin{equation}
|\psi (t)\rangle \rightarrow e^{-F(t)}|\psi (t)\rangle ,\text{ \ }%
H(t)\rightarrow e^{-F(t)}H(t)e^{F(t)},
\end{equation}%
where $F(t)$ is an anti-Hermitian operator and $\psi (t)$ the state of the
system. When $F(t)$ is appropriately chosen to make the first-order term of
the effective Hamiltonian vanishing, i.e., $H_{1}+[H_{0},F]-i\partial _{t}F=0
$, we obtain an effective Hamiltonian to the second order $%
H_{eff}=H_{0}+[H_{1},F]/2$ in principle, and the above equation explicitly
determines $F=F(t)$.

In this section, the canonical transformations are made to obtain the
effective Hamiltonian. In the interaction picture with respect to
\begin{equation*}
H_{0}^{S}=\omega _{z}(S_{z}^{(1)}+S_{z}^{(2)})+\omega
_{z}(B^{+}B+\sum_{k}C_{k}^{+}C_{k}),
\end{equation*}
the Hamiltonian $H^{I}=H_{0}^{I}+H_{1}^{I}+H_{p}^{I}$ contains three parts
\begin{align}
& H_{0}^{I}=\Delta _{1}S_{z}^{(1)}+\Delta _{2}S_{z}^{(2)}\text{,} \\
& H_{1}^{I}=\sum_{i=1,2}f_i\Omega (S_{+}^{(i)}B+S_{-}^{(i)}B^{+})\text{,} \\
& H_{p}^{I}=e^{itH_{0}^{S}}H_{p}^{S}e^{-itH_{0}^{S}}\equiv H_{p}^{S}
\end{align}%
Here,%
\begin{equation*}
\Delta _{j}=\Omega _{z}-\omega _{z}+f_jI_{0}\sum_{i}g_0^{(i)}(j=1,2)
\end{equation*}
are the detunings of electron spin and nuclear spin and hyperfine
interaction.

It can be observed from the Hamiltonian $H^{I}$ that the time-dependent term
$H_{1}^{I}$ can be considered as first-order perturbation with respect to
the zeroth-order term $H_{0}^{I}$ (disregarding $H_{p}^{I}$). Then, we
perform a transformation $\exp (F(t))$ to the Hamiltonian $H^{I}$ to
eliminate the time-dependent term $H_{1}^{I}$, that is, the condition
\begin{equation}
H_{1}^{I}+[H_{0}^{I},F]-i\partial _{t}F=0  \label{condition}
\end{equation}%
should be fulfilled, where TDFT operator is%
\begin{equation*}
F(t)=(x_{1}(t)S_{+}^{(1)}+x_{2}(t)S_{+}^{(2)})B-h.c.
\end{equation*}%
It follows from Eq. (\ref{condition}) that the corresponding coefficients of
$S_{+}^{(1)}B$, $S_{-}^{(1)}B^{+}$, $S_{+}^{(2)}B$ and $S_{-}^{(2)}B^{+}$ at
the left hand side of Eq. (\ref{condition}) vanish, i.e.,
\begin{align}
f_{1}\Omega +\Delta _{1}x_{1}-i\dot{x}_{1}& =0\text{,}  \label{x1} \\
f_{2}\Omega +\Delta _{2}x_{2}-i\dot{x}_{2}& =0\text{.}
\end{align}

In case that the electrons go through the nuclear spins with a uniform speed
$v$, the solutions to the above equations are $x_{j}\simeq -f_j\Omega/\Delta$
($j=1,2$), where $\Delta =\Omega _{z}-\omega _{z}$ is used to replace $%
\Delta _{1}$ and $\Delta _{2}$ since $\Delta \simeq \Delta _{1}\simeq \Delta
_{2}$ in the realistic parameters (for the necessary details please refer to
Appendix \ref{app:appendix2}). Then, the effective Hamiltonian is obtained
approximately as follows
\begin{align}
H^{F}\simeq & H_{0}^{I}+\frac{1}{2}[H_{1}^{I},F]+H_{p}^{I}+[H_{p}^{I},F]
\notag \\
\simeq &[\Delta _{1}-\frac{f_1^2\Omega^2}{\Delta}(2<B^+B>+1)]S_{z}^{(1)}
\notag \\
&+[\Delta _{2}-\frac{f_2^2\Omega^2}{\Delta}(2<B^+B>+1)]S_{z}^{(2)} \\
&-\frac{f_1 f_2\Omega^2 }{\Delta}%
[S_{+}^{(1)}S_{-}^{(2)}+S_{+}^{(2)}S_{-}^{(1)}]+H_{p}^{I}+[H_{p}^{I},F]\text{%
,}  \notag
\end{align}
where $\left\langle B^{+}B\right\rangle $ denotes the average number of
nuclear excitation, and the fast-oscillating terms including the factor $%
exp(\pm i\Delta t)$ have been dropped off.

When almost all nuclear spins are in their ground state, the system is in
the low collective excitation limit, i.e., $\left\langle B^{+}B\right\rangle
\rightarrow 0$. By using%
\begin{equation*}
|\Delta _{1,2}|\simeq |\Delta |\gg f_{1,2}\Omega\text{,}
\end{equation*}
we have
\begin{equation*}
\Delta _{j}-\frac{f_j^2\Omega^2}{\Delta}(2<B^+B>+1) \simeq \Delta _{j}\simeq
\Delta \text{.}
\end{equation*}%
Thus,
\begin{align}
H^{F}\simeq & \Delta (S_{z}^{(1)}+S_{z}^{(2)})-\frac{f_1 f_2\Omega^2}{\Delta}%
(S_{+}^{(1)}S_{-}^{(2)}+h.c.)  \notag \\
& +H_{p}^{I}+[H_{p}^{I},F]\text{.}
\end{align}

In the following calculation, it will be shown that the complex term $%
[H_{p}^{I},F]$ will be dropped in the interaction picture. With respect to $%
H_{0}=\Delta (S_{z}^{(1)}+S_{z}^{(2)})$, the effective interaction
Hamiltonian is
\begin{equation}
H_{eff}=V_{1}+V_{2}\text{,}  \label{Heff}
\end{equation}%
where
\begin{align}
V_{1}& =-e^{iH_{0}t}[\frac{f_1 f_2\Omega^2}{\Delta}%
(S_{+}^{(1)}S_{-}^{(2)}+S_{+}^{(2)}S_{-}^{(1)})]e^{-iH_{0}t}  \notag \\
& =-\frac{f_1 f_2\Omega^2}{\Delta}%
(S_{+}^{(1)}S_{-}^{(2)}+S_{+}^{(2)}S_{-}^{(1)})\text{,}  \label{V1} \\
V_{2}& =e^{itH_{0}}H_{p}^{I}e^{-itH_{0}}=H_p^S \text{.}
\end{align}
In the above calculation, we have dropped the high-frequency terms $\exp
(iH_{0}t)[H_{p}^{I},F]\exp (-iH_{0}t)$ including the factors $\exp (\pm
i\Delta t)$. It is a reasonable approximation which is frequently used in
the Jaynes-Cummings model.

Now, we study the time evolution driven by the above effective Hamiltonian.
First of all, we study a special case that the total system is initially
prepared without the collective excitations of the bus spins. In this case,
the effective interaction $V_{2}$ does not play a role. In a Hilbert space
spanned by the two electron states $\left\vert ee\right\rangle $, $%
\left\vert eg\right\rangle $, $\left\vert ge\right\rangle $ and $\left\vert
gg\right\rangle $, it is clear that there exists an invariant subspace
spanned by $\left\vert eg\right\rangle $ and $\left\vert ge\right\rangle $.
If the system starts from $\left\vert \Psi (0)\right\rangle =\left\vert
eg\right\rangle $, at time $t$ it would definitely evolve into
\begin{equation*}
\left\vert \psi (t)\right\rangle =\cos \theta (t)\left\vert eg\right\rangle
-i\sin \theta (t)\left\vert ge\right\rangle ,
\end{equation*}
where
\begin{equation}
\theta (t)=-\int_{0}^{t}\frac{f_1 f_2\Omega^2}{\Delta}dt^{\prime }\text{.}
\label{theta}
\end{equation}

In comparison with the result in Ref. \cite{Song}, where $H_{p}$ was
considered as a perturbation in the low excitation approximation, we examine
the system evolving under total Hamiltonian containing $V_{1}$ and $V_{2}$.
Then we can get the equations for the coefficients as follow
\begin{align}
i\dot{C}_{ge}& =-\frac{f_{1}f_{2}\Omega ^{2}}{\Delta }%
C_{eg}-(V_{2})_{eg}C_{ge}\text{,} \\
i\dot{C}_{eg}& =(V_{2})_{eg}C_{eg}-\frac{f_{1}f_{2}\Omega ^{2}}{\Delta }%
C_{ge}\text{,}
\end{align}%
where $(V_{2})_{jk}=\langle jk|V_{2}\left\vert jk\right\rangle $ and $%
j,k=e,g $. Similarly, there's an invariant subspace $\{\left\vert
eg\right\rangle \text{,}\left\vert ge\right\rangle \}$.

\section{Spin Entanglement}

In the above sections, we have obtained a typical spin-spin coupling in the
effective Hamiltonian, which is induced by the collective excitations.
Driven by this Hamiltonian, two electron spins can be entangled dynamically.
To characterize the extent of entanglement, we use concurrence to measure
the induced entanglement. For an arbitrary state of two-qubit system
described by the density operator $\rho $, a measure of entanglement can be
defined as the concurrence \cite{Wootters, Wang},
\begin{equation}
C(\rho )=max\{0,\lambda _{1}-\lambda _{2}-\lambda _{3}-\lambda _{4}\}\text{,}
\end{equation}%
where the $\lambda _{i}$'s are the square roots of the eigenvalues of the
non-Hermitian matrix $\rho \widetilde{\rho }$ in decreasing order. And%
\begin{equation*}
\widetilde{\rho }=(\sigma _{y}\otimes \sigma _{y})\rho ^{\ast }(\sigma
_{y}\otimes \sigma _{y}),
\end{equation*}%
where $\rho ^{\ast }$ is the complex conjugate of $\rho $, $\sigma _{y}$ the
Pauli operator. Actually, even \ from the original Hamiltonian (1) we can
also prove that the corresponding reduced density matrix for two spin is of
the form
\begin{equation}
\rho ^{(12)}=\left(
\begin{array}{cccc}
u^{+} & 0 & 0 & 0 \\
0 & w^{1} & z^{\ast } & 0 \\
0 & z & w^{2} & 0 \\
0 & 0 & 0 & u^{-}%
\end{array}
\right) .  \label{reduced}
\end{equation}

To prove the above result, we consider that, in the original Hamiltonian the
interaction terms \
\begin{align}
H_{I}=&
S_{z}^{(1)}\sum_{j=1}^{N}g_{1}^{(j)}I_{z}^{(j)}+S_{z}^{(2)}%
\sum_{j=1}^{N}g_{2}^{(j)}I_{z}^{(j)}  \notag \\
& +S_{+}^{(1)}\sum_{l=1}^{N}\frac{g_{1}^{(j)}}{2}I_{-}^{(j)}+S_{+}^{(2)}%
\sum_{j=1}^{N}\frac{g_{2}^{(j)}}{2}I_{-}^{(j)}+h.c.\text{,}
\end{align}%
conserves the total spin z-component
\begin{equation*}
\mathcal{S}_{z}=\sum_{j=1}^{N}I_{z}^{(j)}+S_{z}^{(1)}+S_{z}^{(2)}
\end{equation*}%
i.e., $[H_{I},\mathcal{S}_{z}]=0$. For such conserved system we express the
concurrence characterizing quantum entanglement in terms of observables,
such as correlation functions.

The complete basis vectors of the total system are denoted by
\begin{eqnarray}
|S_{1},S_{2},\{I_{j}\}\rangle &=&|S_{1},S_{2};I_{1},..,I_{N}\rangle  \notag
\\
&=&\prod\limits_{j=1}^{N}|I_{j}\rangle \otimes |S_{1}\rangle \otimes
|S_{2}\rangle
\end{eqnarray}%
where $|I_{j}\rangle $ is nuclear spin state and $|S_{1,2}\rangle $ denote
the electronic spins$\ \ (I_{j},S_{1,2}=0,1)$ respectively. The fact that $%
\mathcal{S}_{z}$ is conserved can be reflected by the vanishing of some
matrix elements of the density operator $\rho =\rho (H)$ on the above basis
for any state of the total system, that is,

\begin{equation}
\rho _{S_{1},S_{2};I_{1},..,I_{N}}^{S_{1}^{\prime },S_{2}^{\prime
};I_{1}^{\prime },..,I_{N}^{\prime }}=\rho
_{\{n_{j},s_{j}\}}^{\{n_{j},s_{j}\}}\delta (s,s^{\prime },I,I^{\prime }),
\end{equation}%
where
\begin{equation*}
\delta (s,s^{\prime },I,I^{\prime })=\delta (S_{1}+S_{2}-S_{1}^{\prime
}-S_{2}^{\prime }+\sum_{j=1}^{N}(I_{j}-I_{j}^{\prime }))
\end{equation*}

The functional $\rho (H)$ of the Hamiltonian may be a ground state or
thermal equilibrium states. The reduced density matrix $\rho
^{(12)}=Tr_{I}[\rho (H)]$ for two spins, e.g., $S_{1}$ and $%
S_{2}$ are obtained as
\begin{equation}
\lbrack \rho ^{(12)}]_{S_{1},S_{2};}^{S_{1}^{\prime },S_{2}^{\prime
};}=\delta (s,s^{\prime },0,0)\sum_{\{I_{j}\}}\rho
_{S_{1},S_{2};I_{1},..,I_{N}}^{S_{1}^{\prime },S_{2}^{\prime
};I_{1},..,I_{N}}
\end{equation}%
by tracing over all nuclear variables. The corresponding reduced
density matrix for two spins $1$ and $2$ is of the form in
Eq.(\ref{reduced}).
Using the observable quantities, the quantum correlation%
\begin{eqnarray}
z &=&\left\langle \psi \right\vert S_{1}^{+}S_{2}^{-}\left\vert \psi
\right\rangle , \\
u^{\pm } &=&\left\langle \psi \right\vert \left( 1/2\pm S_{1}^{z}\right)
\left( 1/2\pm S_{2}^{z}\right) \left\vert \psi \right\rangle ,  \notag
\end{eqnarray}%
the\ concurrence is rewritten as a computable form
\begin{equation}
C_{12}=2\max (0,\left\vert z\right\vert -\sqrt{u^{+}u^{-}}).
\label{c2}
\end{equation}%
We note that this formula for the concurrence of two electron spins in the
coupled system is the same as that for a spin-1/2 coupling system modeled by
the effective Hamiltonian \cite{Wootters, Wang}. This general form is
consistent with that obtained straightforwardly from the effective
Hamiltonian given in the last section.

\begin{figure}[ptb]
\includegraphics[bb=104 281 493 558,width=7.5 cm]{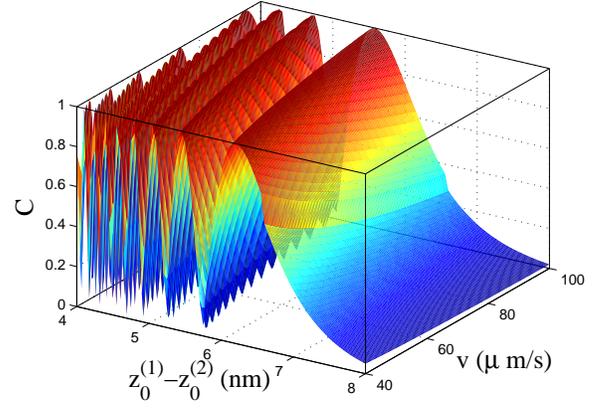}
\caption{(Color Online) The relation between concurrence $C$ and velocity $v$%
, distance $z_0^{(1)}-z_0^{(2)}$, on condition that $a=4$ nm, $%
\Delta=8\times10^{11}$ Hz, $I_{0}=3/2$, $n=0$. }
\label{V1_total}
\end{figure}

\begin{figure}[ptb]
\includegraphics[bb=104 260 473 567,width=7 cm]{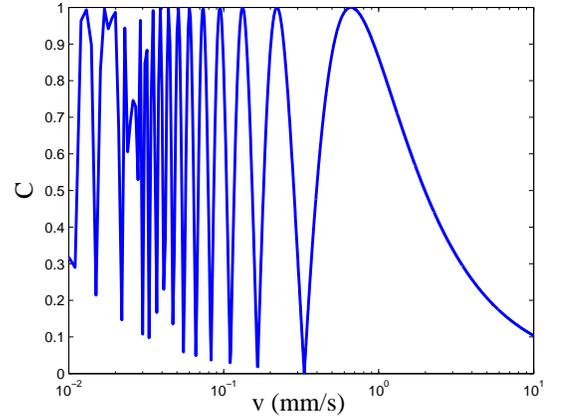}
\caption{Cross section of Fig.\protect\ref{V1_total} with $%
z_0^{(1)}-z_0^{(2)}=4$ nm.}
\label{V1_sectional}
\end{figure}

With the above general consideration, we now study quantitatively the
concurrence for the quantum entanglement of the two electrons passing the
nuclear spins at a uniform speed $v$. In Fig. \ref{V1_total}, the
concurrence is plotted while the speed $v$ and the initial distance between
the two electrons $z_{0}^{(1)}-z_{0}^{(2)}$ are varied. It is obvious that
the concurrence fluctuates from $0$ to $1$ in the low speed region (see also
Fig. \ref{V1_sectional}). When the electrons move with a relative low speed,
the concurrence oscillates rapidly since a lower speed means more time for
evolution from a direct product state towards an entangled state. As the
speed increases, the concurrence falls monotonously if it is bigger than a
certain value. According to Eq. (\ref{theta}), the maximum entangled state
can be obtained when
\begin{equation}
\theta =(n+1)\frac{\pi }{4}\text{.}
\end{equation}%
However, the general relation between the concurrence and $%
z_{0}^{(1)}-z_{0}^{(2)}$ is a little more complicated. The further the two
electrons separate from each other, the longer time both of them need to
pass through the nuclear ensemble. On the other hand, the matrix elements of
the effective interaction $V_{1}$ in Eq. (\ref{V1}), i.e., $\langle \uparrow
\downarrow |V_{1}|\downarrow \uparrow \rangle $, h.c., drop dramatically as
the inter-spin distance increases. This observation is obviously correct
from an intuitively physical consideration. For a longer inter-spin
distance, the spatial wave function of two spins has a smaller overlap, and
then the effective coupling is weak.

\begin{figure}[ptb]
\includegraphics[bb=103 262 484 565,width=7.5 cm]{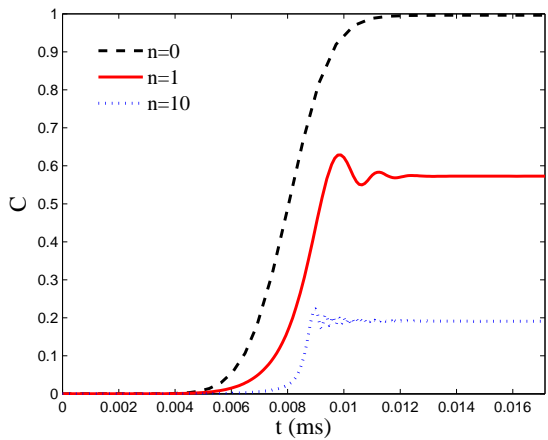}
\caption{(Color Online) The evolution of concurrence $C$ on condition that $%
a=4$ nm, $\Delta=8\times10^{11}$ Hz, $I_{0}=3/2$, $v=0.7$ mm/s.}
\label{V1_V1+V2}
\end{figure}

In the last section, we have only taken $V_{1}$ into consideration. However,
due to the nuclear excitation, i.e., $n>0$, $V_{2}$ will lead to
decoherence. According to Ref. \cite{Song}, under the quasihomogeneous
condition, we have $\sum_{i}g_{0}^{(i)}(I_{z}^{(i)}+I_{0})=n\overline{g}$,
where $\overline{g}=\sum_{i}g_{0}^{(i)}/N$, $n=\langle B^{+}B\rangle $ is
the average number of collective excitation. Thus, the single particle
perturbation can be approximated as
\begin{equation}
V_{2}=H_{p}^{S}\simeq n\overline{g}(f_{1}S_{z}^{(1)}+f_{2}S_{z}^{(2)})\text{.%
}
\end{equation}%
In Fig. \ref{V1_V1+V2}, we plot the concurrence evolution under $%
H_{eff}=V_{1}+V_{2}$. As shown in the figure, the two spins evolve into a
maximum entangled state with appropriate parameters when $n=0$. On the
contrary, the concurrence is suppressed when there exists collective
excitation in the nuclei. Moreover, as more nuclei are excited, the
concurrence falls dramatically.

\section{Conclusion}

In summary, we have proposed a scheme to entangle two electron spins via an
ensemble of nuclei. We also explore the influence of its collective
excitation on the concurrence characterizing two spin entanglement.
Theoretically, the maximum entangled state can be obtained if the electrons
move in a suitable way. Furthermore, with the optimized experimental
parameters, the operation time is within the relaxation time of electron
spins in solid state systems, i.e., the order of ms \cite{Elzerman}.

However, this scheme may encounter some difficulties in practice since it is
based on the low excitation requirement of nuclear ensemble. In recent
experiments, the number of excitation $n$ is of the order of $60\% N$ \cite%
{Bracker}. Thus, further progress in experiment, i.e., lowering the
temperature or optical excitation, are expected to prepare all
nuclear spins in their ground states in order to put this scheme
into practice. Actually, there are only the collective excitations
considered as the quantum data bus to coherently link two spins so
that the inter-spin entanglement is induced. If there exists
noncollective excitations, then extra decoherence will be induced to
break our scheme presented in this paper. Further investigations are
needed for these questions. However, if we can cool the nuclear
ensemble via some new mechanism, e.g., similar to
Ref.\cite{Zhang,Coolexpts}, our scheme will probably work well.

\section*{Acknowledgement}

Thanks very much for helpful discussion with Peng Zhang, Nan Zhao, Zhangqi
Yin, Zhensheng Dai, Yansong Li. This work is supported partially by the 973
Program Grant Nos. 2006CB921106, 2006CB921206 and 2005CB724508, National
Natural Science Foundation of China, Grant Nos. 10325521, 60433050,
60635040, 10474104, and 90503003.

\appendix

\section{Derivation of $g_{1}^{(i)}$ and $g_{2}^{(i)}$}

\label{app:appendix1}

According to Ref.\cite{Schliemann}, the hyperfine interaction constant is
mainly proportional to the electron spin density located at the nucleus.
Thus,
\begin{equation}
g_{1}^{(i)}=\frac{4\mu_{0}}{3I_0\hbar}\mu_B \mu_I\left\vert
\psi_{1}^{(i)}\right\vert ^{2}\text{,}
\end{equation}
where $\mu_{0}$ is vacuum permeability, $I_0$ total nuclear spin quantum
number, $\mu_B$ the Bohr magneton, $\mu_I$ the nuclear magnetic moment, $%
\psi_{1}^{(i)}$ the wavefunction for electron 1 located at the $i$'th
nuclear spin. In a semiconductor crystal, the wavefunction is given by the
product of the Bloch amplitude $u(\vec{r})$ and an envelope function $\Psi(%
\vec{r})$, i.e., $\psi(\vec{r})=u(\vec{r})\Psi(\vec{r})$. In a realistic
crystal $\eta=\left\vert u(\vec{r})\right\vert ^{2}$ reaches a climax at the
lattice positions. It was discovered that $\eta_{As}=4.5\times10^3$ and $%
\eta_{Ga}=2.7\times10^3$ \cite{Paget}. On account of isotope abundance and
their different nuclear magnetic moment \cite{AIPH}, $\overline{\mu_I\eta}%
=3.153\times10^{-23}A\cdot m^2$. In our simulation, we assume that the
wavefunction is Gaussian wave packet with initial location $z_i^{(1)}$ and a
uniform speed $v$, that is
\begin{equation}
\left\vert \psi_{1}^{(i)}\right\vert ^{2}=\left(\frac{4}{\pi a^{2}}%
\right)^{3/2}e^{-4[x_i^2+y_i^2+(z_0^{(1)}+vt-z_i)^2 ]/a^2}\text{,}
\end{equation}
where $\vec{r}_{i}=x_i\hat{x}+y_i\hat{y}+z_i\hat{z}$ is the position of the $%
i$'th nuclear spin with respect to the center of nuclei.

Since the nuclei are distributed in a flat box with $\left\vert
x_i\right\vert\leq a$, $\left\vert y_i\right\vert\leq a$, $\left\vert
z_i\right\vert\leq a/5$. Therefore,
\begin{align}
g_{1}^{(i)}&\simeq\frac{4\mu_{0}}{3I_0\hbar}\mu_B \mu_I\eta\left(\frac{4}{%
\pi a^{2}}\right)^{3/2}e^{-4r_i^2/a^2}e^{-4(z_0^{(1)}+vt)^2/a^2}  \notag \\
&= g_{0}^{(i)}f_{1}(t)\text{,}
\end{align}%
where
\begin{align}
g_{0}^{(i)}&=\frac{4\mu_{0}}{3I_0\hbar}\mu_B \mu_I\eta\left(\frac{4}{\pi
a^{2}}\right)^{3/2}e^{-4r_i^2/a^2}\text{,} \\
f_{1}(t)&=e^{-4(z_0^{(1)}+vt)^2/a^2}\text{.}
\end{align}
Here, we have neglected the term $e^{-8(z_0^{(1)}+vt)z_i/a^2}$ based
on the following consideration. On the one hand, the above
approximation tends to be exact as the quasi-2D quantum well becomes
narrower in the z-direction, e.g., $\left\vert
z_i\right\vert\rightarrow 0$. On the other hand, the effective
coupling intensity is increased as more nuclear spins are included
when $\left\vert z_i\right\vert$ get larger. Thus, optimal value is
chosen for the valid approximation. Similarly, we have
\begin{equation}
g_{2}^{(i)}\simeq g_{0}^{(i)}f_{2}(t)\text{,}
\end{equation}
where%
\begin{equation}
f_{2}(t)=e^{-4(z_0^{(2)}+vt)^2/a^2}\text{.}
\end{equation}

\section{Derivation of $x_{i}$}

\label{app:appendix2}

According to Eq.(\ref{x1}), one has
\begin{align}
x_{1} &=-i\Omega e^{-i\int_{0}^{t}\Delta _{1}dt^{\prime
}}\int_{0}^{t}f_{1}e^{i\int_{0}^{t^{\prime }}\Delta _{1}dt^{\prime \prime
}}dt^{\prime }  \notag \\
&\simeq -i\Omega e^{-i\Delta t}\int_{-\infty }^{t}f_{1}e^{i\Delta t^{\prime
}}dt^{\prime }  \notag \\
&=\frac{-\Omega }{\Delta }e^{-i\Delta t}[f_{1}e^{i\Delta t}-\frac{8v}{a^{2}}%
\int_{-\infty }^{t}f_{1}e^{i\Delta t^{\prime }}(z_{0}^{(1)}+vt^{\prime
})dt^{\prime }]  \notag \\
&\simeq \frac{-\Omega }{\Delta }e^{-i\Delta t}(f_{1}e^{i\Delta t}-\frac{8v}{a%
}\int_{-\infty }^{t}f_{1}e^{i\Delta t^{\prime }}dt^{\prime })\text{.}  \notag
\end{align}
Here, we have replaced $\Delta _{1}$ by $\Delta =\Omega _{z}-\omega _{z}$
since%
\begin{equation*}
f_1I_{0}\sum_i g_0^{(i)}\lesssim 8.318\times10^{10}\text{ Hz}\ll \Delta
=8\times10^{11}\text{ Hz.}
\end{equation*}
Furthermore, we have replaced
\begin{equation*}
\frac{8v}{a^{2}}\int_{-\infty }^{t}f_{1}e^{i\Delta t^{\prime
}}[z_{0}^{(1)}+vt^{\prime }]dt^{\prime }
\end{equation*}
by
\begin{equation*}
\frac{8av}{a^{2}}\int_{-\infty }^{t}f_{1}e^{i\Delta t^{\prime }}dt^{\prime }%
\text{,}
\end{equation*}%
since one notices the fact that $z_{0}^{(1)}+vt^{\prime }\sim a$ is the
effective integration range and the change of $f_{1}(t^{\prime })$ (also
change of $vt^{\prime }$) is much slower than that of $e^{i\Delta t^{\prime
}}$. Thus,
\begin{eqnarray*}
x_{1} &\simeq \frac{-\Omega }{\Delta }f_{1}-i\frac{8v}{\Delta a}x_{1}\text{,}
\end{eqnarray*}
Generally speaking, $\Delta $ depends on the applied magnetic field. In case
that $\Delta =8\times10^{11}$ Hz and $a\thicksim 4$nm, we have $%
8v/a\Delta\ll 1 $ for all $v\ll 400$ m/s. Thus, we have
\begin{eqnarray}
x_{1} &\simeq \frac{-\Omega }{\Delta }f_{1}\text{.}
\end{eqnarray}
Similarly,
\begin{eqnarray}
x_{2} &\simeq \frac{-\Omega }{\Delta }f_{2}\text{.}
\end{eqnarray}

\end{document}